\documentclass[preprint,10pt,3p]{elsarticle}

\usepackage{hyperref}

\usepackage{amsmath}
\usepackage{amsfonts}
\usepackage{euscript}
\usepackage{algorithm2e}
\usepackage{color}

\journal{Information Processing Letters}

\newtheorem{thm}{Theorem}
\newtheorem{lem}[thm]{Lemma}
\newtheorem{cor}[thm]{Corollary}
\newdefinition{dfn}{Definition}
\newproof{pf}{Proof}

\DeclareMathOperator*{\E}{\textrm{E}}
\SetKwFunction{OSW}{OSW}

\begin{document}

\begin{frontmatter}

\title{Small World Model based on a Sphere Homeomorphic Geometry}

\author{Santiago Viertel}
\ead{sviertel@inf.ufpr.br}
\author{Andr\'e Lu\'is Vignatti}
\ead{vignatti@inf.ufpr.br}
\address{DINF, Federal University of Paran\'a, Curitiba, Brazil}

\begin{abstract}

We define a small world model over the octahedron surface and relate its distances with those of embedded spheres, preserving constant bounded distortions.
The model builds networks with both number of vertices and size $\Theta\left(n^2\right)$, where $n$ is the size parameter.
It generates long-range edges with probability proportional to the inverse square of the distance between the vertices. 
We show a greedy routing algorithm that finds paths in the small world network with $\mathcal{O}\left(\log^2n\right)$ expected size.
The probability of creating cycles of size three (C3) with long-range edges in a vertex is $\mathcal{O}\left(\log^{-1}n\right)$.
Furthermore, there are $\Theta\left(n^2\right)$ expected number of C3's in the entire network.

\end{abstract}


\end{frontmatter}


\section{Introduction}
\label{sec:introduction}

Stanley Milgram~\cite{milgram1967small} conclude that social networks have a large number of paths with small length.
This motivates the proposal of small world graphs  models~\cite{watts1998cds,Kleinberg:2000,Kleinberg:2001:SPD:2980539.2980596,Liben-Nowell16082005}.
Kleinberg~\cite{Kleinberg:2000} presents a model that generates a $n\times n$ lattice of vertices $V=\{1,2,...,n\}\times\{1,2,...,n\}$.
He defines the \emph{lattice distance} between two vertices $(i,j),(k,l)\in V$ as $d((i,j),(k,l))=|k-i|+|l-j|$.
The model has three parameters, $p\ge1$, $q\ge0$ and $r\ge0$.
It links each vertex with directed edges to the vertices within lattice distance $p$, and for each $u\in V$, it generates $q$ directed edges (independent random trials) to $v\in V\setminus\{u\}$ with  probability proportional to $d^{-r}(u,v)$.
The probability of each $v$ is multiplied by the \emph{normalizing factor} $\left(\sum_{w\in V\setminus\{u\}}d^{-r}(u,w)\right)^{-1}$.
Kleinberg calls it as the \emph{inverse $r^\textrm{th}$-power distribution}.
We call \emph{long-range edges} those random generated edges. Kleinberg proves that, for $p=q=1$ and $r=2$, there is a greedy routing algorithm that finds paths with $\mathcal{O}\left(\log^2n\right)$ expected length. For each vertex, the algorithm takes constant time and logarithmic space (in bits).
Small world networks has routing applications in P2P networks~\cite{AspnesDS2002,ZHANG2004555,Manku:2004:KTN:1007352.1007368}, MANETs~\cite{4067680} and WSN~\cite{4678803,Liu2009}.
Most uses the ideas of Kleinberg, generating a clustered network and long-range edges with the inverse $r^\textrm{th}$-power distribution.

Manku, Naor and Wieder~\cite{Manku:2004:KTN:1007352.1007368} present a greedy routing algorithm that considers the vertex neighbors and the neighbors of neighbors in a routing decision.
The algorithm finds paths with $\mathcal{O}(\log n/\log\log n)$ expected length in a $n$-ring with $\log n$ long-range edges per vertex.
Martel and Nguyen~\cite{Martel:2004:AKS:1011767.1011794} extend the Kleinberg model to the $d$-dimensional lattice.
The greedy routing algorithm finds paths with $\mathcal{O}\left(\log^{1+1/d}n\right)$ expected length for $r=d$.
In this algorithm, the vertices have positioning information of the long-range edges of the closest $\log n$ neighbors.
Fraigniaud, Gavoille and Paul~\cite{Fraigniaud:2006:ESE:1160297.1160302} present similar results with an oblivious algorithm. Zeng, Hsu and Wang~\cite{Zeng:2005:NOR:2098796.2098861} present a model over a unidirectional $n$-ring.
Each vertex has one long-range edge generated with the inverse $1^\textrm{th}$-power distribution and two augmented local edges to vertices within $\log^2 n$ lattice distance chosen uniformly at random.
They present two algorithms that find paths with $\mathcal{O}(\log n\log\log n)$ expected length.
One year later, the first two authors~\cite{Zeng2006} generalize the model to dimensions $d$ and define a routing algorithm that find paths with $\mathcal{O}(\log n)$ expected length.
Liu, Guan, Bai and Lu~\cite{Liu2009} present a model that generates a $n\times n$ lattice subdivided in $k\times k$ clusters.
Only one vertex of each cluster has a long-range edge to another vertex of other cluster generated with the inverse $2^\textrm{th}$-power distribution.
The routing algorithm finds paths with $\mathcal{O}(\log m\log^2\log m)$ expected length, where $m=n/k$ and each vertex has positioning information of the $\mathcal{O}(\log m)$ closest long-range edges.

Some works~\cite{Manku:2004:KTN:1007352.1007368,Martel:2004:AKS:1011767.1011794,Zeng:2005:NOR:2098796.2098861,Liu2009} present models that build a base graph over bounded geometries that are plane in the three-dimensional euclidean space.
Kleinberg's model~\cite{Kleinberg:2000} is an example that generates a clustered network over the square with vertices $(1,1,0)$, $(1,n,0)$, $(n,n,0)$ and $(n,1,0)$.
Other work~\cite{Martel:2004:AKS:1011767.1011794} presents a model that use the two-dimensional torus, which is a solid geometry with genus one.
There is a lack of small world models that generate the base graph over genus-zero solid geometries, that is, are homeomorphic to spheres, such as, for example, any Platonic solid.
Our octahedral small world model (\OSW) generates the base graph over the octahedron (Section~\ref{sec:octahedralSmallWorldModel}).
We relate the base graph with spheres (Section~\ref{subsec:theNOctahedralGraphAndSpheres}), providing a connection with global friendship networks. The vertices on the octahedron can be projected on the surface of spheres.
We relate the distances between neighbors vertices on both and the size $n$ of the octahedron with spheres radii.
As in the octahedron, where the distances are bounded by a constant for increasing $n$, there are spheres where the distances are also bounded by a constant for increasing radius.
Then, there is a family of spheres that are asymptotically similar to the octahedron in terms of distances on the surface.
We also define a greedy routing algorithm that finds paths of size $\mathcal{O}\left(\log^2n\right)$ in octahedral small world graphs.
Moreover, the expected number of C3's in these graphs is $\Theta\left(n^2\right)$ (Section~\ref{sec:c3NotInTheNOctahedralGraph}).
Theorems \ref{thm:Eu} and \ref{thm:expectedNumberOfC3sNotInTheNOctahedralGraph} can be used, for example, in running time analysis of algorithms that perform searches of C3's.
Identifying the base graph performing a local search of cycles in each vertex is an approach to label vertices~\cite{2018arXiv180601469V}.
\section{Octahedrons and Graphs}
\label{sec:octahedronsAndGraphs}

\begin{figure}
	\centering
	\begin{minipage}[b]{.44\textwidth}
		\centering
		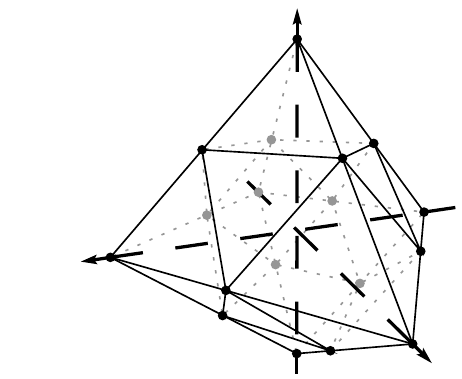
		\caption{A two-octahedral graph.}
		\label{fig:2OctahedralGraph}
	\end{minipage}
	\qquad
	\begin{minipage}[b]{.5\textwidth}
		\centering
		\includegraphics{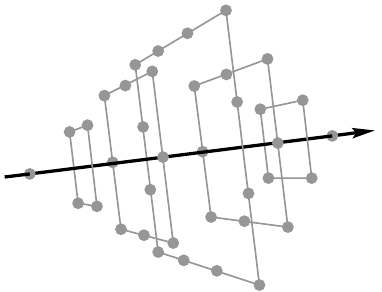}
		\caption{The underlying cycles surrounding an axis.}
		\label{fig:UnderlyingCycles}
	\end{minipage}
\end{figure}

Let the \emph{$r$-octahedron} be the set $O_r=\left\{u\in\mathbb{R}^3:\sum_{i=1}^3|u_i|=r\right\}$, with $r>0$ and $u=(u_1,u_2,u_3)$.
Let $V=O_n\cap\mathbb{Z}^3$ for $n\in\{1,2,3,...\}$, i.e., $V$ is the set of all 3D vectors with integer coordinates in the $n$-octahedron.
We generate an undirected graph on $V$ that ``wraps'' $O_n$.
The edges do not pass through the ``inside'' of $O_n$, linking only the closest vectors on $O_n$. So, each $v\in V$ has an undirected edge to all $w\in V\setminus\{v\}$ such that $|v_i-w_i|\le1$, for $1\le i\le3$.
Definition~\ref{dfn:nOctahedralGraph} formalizes the $n$-octahedral graph.
Figure~\ref{fig:2OctahedralGraph} illustrates a two-octahedral graph.
Lemma~\ref{lem:numberOfVertices} shows the number of vertices of $G'_n$.
Note that, $n$ is the size of the $n$-octahedral graph and $|V|$ is the number of vertices.
Lemma~\ref{lem:numberOfEdges} shows the number of edges of $G'_n$.

\begin{dfn}
	The \emph{$n$-octahedral graph} with size $n\in\{1,2,3,...\}$ is $G'_n=(V,E')$ such that $V=\{u\in\mathbb{Z}^3:\sum_{i=1}^3|u_i|=n\}$ and $E'=\{\{v,w\}\subset V:|v_i-w_i|\le1, \forall 1\le i\le3\}$.
	\label{dfn:nOctahedralGraph}
\end{dfn}

\begin{lem}
	$|V|=4n^2+2$.
	\label{lem:numberOfVertices}
\end{lem}
\begin{pf}
	We count all combinations of $u_1$, $u_2$ and $u_3$ that satisfy the Definition~\ref{dfn:nOctahedralGraph}.
	For $u_1=-n$ and $u_1=n$, $u_2=u_3=0$, that are two combinations.
	For each $-n+1\le u_1\le n-1$, for $u_2=-(n-|u_1|)$ and $u_2=n-|u_1|$, $u_3=0$, that are two combinations for each $u_1$.
	For each $-n+1\le u_1\le n-1$ and $-(n-|u_1|)+1\le u_2\le(n-|u_1|)-1$, $u_3=-(n-|u_1|-|u_2|)$ and $u_3=n-|u_1|-|u_2|$, that are two combinations for each $u_1$ and $u_2$.
	Then, $$|V| = 2 + \left(\sum_{u_1=-n+1}^{n-1}2\right) + \left(\sum_{u_1=-n+1}^{n-1}\left(\sum_{u_2=-(n-|u_1|)+1}^{(n-|u_1|)-1}2\right)\right)=4n^2+2.$$\qed
\end{pf}

\begin{lem}
	$|E'|=12n^2$.
	\label{lem:numberOfEdges}
\end{lem}
\begin{pf}
	By Definition~\ref{dfn:nOctahedralGraph}, for all $u\in V$, the two neighbors of $u$ with the same value of the third coordinate, when $-n+1\le u_3\le n-1$, are:
	\begin{itemize}
		\item $(-1,u_2-1,u_3)\in V$ and $(1,u_2-1,u_3)\in V$ if $u_1=0$ and $u_2>0$;
		\item $(-1,u_2+1,u_3)\in V$ and $(1,u_2+1,u_3)\in V$ if $u_1=0$ and $u_2<0$;
		\item $(u_1-1,-1,u_3)\in V$ and $(u_1-1,1,u_3)\in V$ if $u_1>0$ and $u_2=0$;
		\item $(u_1+1,-1,u_3)\in V$ and $(u_1+1,1,u_3)\in V$ if $u_1<0$ and $u_2=0$;
		\item $(u_1-1,u_2+1,u_3)\in V$ and $(u_1+1,u_2-1,u_3)\in V$ if $u_1>0$ and $u_2>0$ or $u_1<0$ and $u_2<0$;
		\item $(u_1-1,u_2-1,u_3)\in V$ and $(u_1+1,u_2+1,u_3)\in V$ if $u_1<0$ and $u_2>0$ or $u_1>0$ and $u_2<0$.
	\end{itemize}
	This arrangement of edges defines a sequence of cycles with increasing sizes for $u_3$ growing from $-n+1$ to zero and decreasing sizes from one to $n-1$, as Figure~\ref{fig:UnderlyingCycles} shows.
	There are two similar sequences of cycles for $u_1$ and $u_2$.
	Counting one edge for each vertex of each cycle in each coordinate, $$|E'|=\sum_{\{i,j\}\in\binom{\{1,2,3\}}{2}}
	\left(\left(\sum_{u_i=-n+1}^{n-1}2\right)+\left(\sum_{u_i=-n+1}^{n-1}\left(\sum_{u_j=-(n-|u_i|)+1}^{(n-|u_i|)-1}2\right)\right)\right)=12n^2.$$\qed
\end{pf}

\subsection{The n-Octahedral Graph and Spheres}
\label{subsec:theNOctahedralGraphAndSpheres}

\begin{figure}
	\centering
	\begin{minipage}[b]{.24\textwidth}
		\centering
		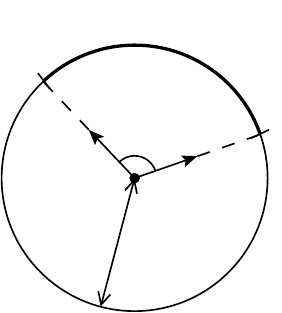
		\caption{Distance between the projections of $u$ and $v$, $u'$ and $v'$ respectively, on a sphere with radius $r>0$.}
		\label{fig:Projections}
	\end{minipage}
	\quad
	\begin{minipage}[b]{.29\textwidth}
		\centering
		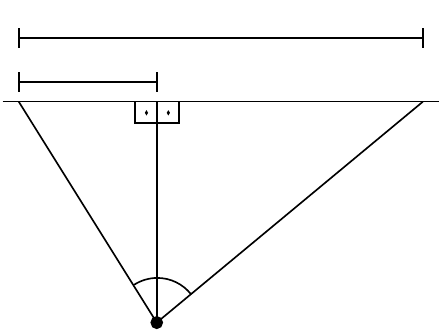
		\caption{Right triangles defined by $u$, $v$ and $O_n$.}
		\label{fig:MaximumAngle}
	\end{minipage}
	\quad
	\begin{minipage}[b]{.39\textwidth}
		\centering
		\includegraphics{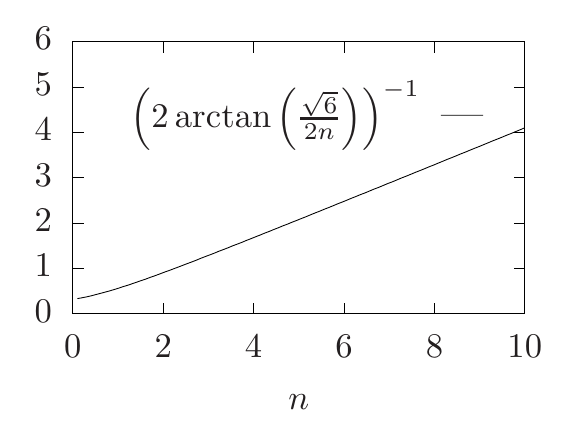}
		\caption{The upper bound function of $r>0$.}
		\label{fig:InternalSphereRadius}
	\end{minipage}
\end{figure}

Let the distance between a pair $\{u,v\}\in E'$ on $O_n$ be the length in the euclidean space of the line segment bounded by $u\in V$ and $v\in V$.
Let the distance between $u$ and $v$ on the sphere with radius $r>0$ be the length in the euclidean space of the smallest circular arc defined by the projections of $u$ and $v$ on the surface of the sphere with radius $r$ and center at $(0,0,0)$.
Figure~\ref{fig:Projections} shows the projections of $u$ and $v$, the circular arc defined by them and the angle $\alpha_{uv}$ between $u$ and $v$ in radians.
The \emph{great-circle distance} is the well-known relation $\frac{d^r_{uv}}{2\pi r}=\frac{\alpha_{uv}}{2\pi}$.
Thus $d^r_{uv}=r\cdot\alpha_{uv}$ for all $\{u,v\}\in E'$, where $d^r_{uv}$ is the distance between $u$ and $v$ on the sphere with radius $r$.

By Definition~\ref{dfn:nOctahedralGraph}, the distance on $O_n$ between $u$ and $v$ is $\sqrt{2}$ for all $\{u,v\}\in E'$ and $n\ge1$.
On the other hand, the distances on the sphere $d^r_{uv}$ vary according to $\alpha_{uv}$ for all radius $r>0$.
In the case of $r=n$, $d^n_{uv}$ diverge as $n$ tends to infinity, because $r$ increases faster than $\alpha_{uv}$ inversely decrease.
On the other extreme, when $r$ is a positive constant $\lambda'$, $d^{\lambda'}_{uv}$ converge to $0$ as $n$ tends to infinity.
Theorem~\ref{thm:upperBoundForSphereRadius} shows that all $d^r_{uv}$ are in the interval $(0,\lambda]$ when $r\le\lambda\left(2\arctan\left(\frac{\sqrt{6}}{2n}\right)\right)^{-1}$ for a positive constant $\lambda$.
Therefore, the distances on the sphere are $\mathcal{O}(1)$ for some values of $r$, which is asymptotically similar to the distances on $O_n$.
Figure~\ref{fig:InternalSphereRadius} shows the radius upper bound function plot for $\lambda=1$.

\begin{thm}
	The distances on all spheres with radius $0<r\le\lambda\left(2\arctan\left(\frac{\sqrt{6}}{2n}\right)\right)^{-1}$ are $0<d^r_{uv}\le\lambda$, for all $\{u,v\}\in E'$, $n\ge1$ and a constant $\lambda>0$.
	\label{thm:upperBoundForSphereRadius}
\end{thm}
\begin{pf}
	For $d^r_{uv}\le\lambda$ for all $\{u,v\}\in E'$, $r\le\lambda\cdot\alpha_{uv}^{-1}$.
	The analysis follows on all planes defined by all $u$, $v$ and $(0,0,0)$ as shown in Figure~\ref{fig:MaximumAngle}.
	Given the legs $a$ and $b$ of the two right triangles defined in each plane, $\alpha_{uv}=\arctan(a/b)+\arctan\left(\left(\sqrt{2}-a\right)/b\right)$.
	The values of all $\alpha_{uv}$ are maximum for $\frac{\partial\alpha_{uv}}{\partial a}=0$, that is $a=\sqrt{2}/2$.
	Moreover, $\alpha_{uv}$ is upper bounded when $b$ is minimum because $a$ is constant, for all $u$ and $v$.
	The minimum value of $b$ is the radius of the inscribed sphere in $O_n$, that is $\left(\sqrt{3}/3\right)n$.
	Then $\alpha_{uv}\le2\arctan\left(\frac{\sqrt{6}}{2n}\right)$ for all $u$ and $v$ and $r\le\lambda\left(2\arctan\left(\frac{\sqrt{6}}{2n}\right)\right)^{-1}\le\lambda\cdot\alpha_{uv}^{-1}$.\qed
\end{pf}
\section{Octahedral Small World Model}
\label{sec:octahedralSmallWorldModel}

Let $d:V\times V\rightarrow\mathbb{N}$ be the distance function defined by the length of a minimum path between all pairs of vertices $u,v\in V$ in $G'_n$.
We define \emph{path} as a sequence of distinct vertices which each two consecutive vertices the first is incident to the second, \emph{length} of a path as the number of vertices of the path minus one and \emph{minimum path} as a path with the minimum length over all paths.
Let $C_{uv}$ be the event of $u\in V$ choosing $v\in V\setminus\{u\}$ to create the directed edge $(u,v)\in E$ and $Z_u=\left(\sum_{w\in V\setminus\{u\}}d_{uw}^{-2}\right)^{-1}$ be the \emph{normalizing factor} of all $u$.

\begin{dfn}[\OSW model]
	The \emph{octahedral small world} (\OSW) model is $G_n=(V,E)$ such that for the $n$-octahedral graph $G'_n=(V,E')$:
	\begin{enumerate}[(i)]
		\item for all $\{u,v\}\in E'$, both directed edges $(u,v)$ and $(v,u)$ are included in $E$ and;
		\item for all $u\in V$ and a $v\in V\setminus\{u\}$, $(u,v)$ is included in $E$ with probability $\Pr(C_{uv})=Z_u\cdot d_{uv}^{-2}$.
	\end{enumerate}
	\label{dfn:OSW}
\end{dfn}

Note that $G'_n$ is undirected, as shown in Definition~\ref{dfn:nOctahedralGraph}, and $G_n$ is directed, as shown in Definition~\ref{dfn:OSW}.
All undirected edges $\{u,v\}\in E'$ correspond to the pair of directed edges $(u,v),(v,u)\in E$.
We call \emph{long-range edges} those generated by the independent random trials of \OSW.
Next, we design a greedy routing algorithm that finds small paths in $G_n$.
Given a vertex $u\in V$ and a message, the algorithm sends the message to the vertex $v\in\EuScript{N}^+_{G_n}(u)$ with minimum angle $\alpha_{vt}$ with the target vertex $t\in V$, where $\EuScript{N}^+_{G_n}(u)$ is the set of out-neighbors of $u$ in $G_n$.
The algorithm selects $v$ computing argmax$_{v\in\EuScript{N}^+_{G_n}(u)}\left(\frac{v\cdot t}{|v||t|}\right)$, where $v\cdot t=\sum_{i=1}^3v_it_i$, $|w|=\left(\sum_{i=1}^3w_i^2\right)^{1/2}$ and $t$ is in the message header.

\begin{figure}
	\centering
	\begin{minipage}[b]{.36\textwidth}
		\centering
		\includegraphics{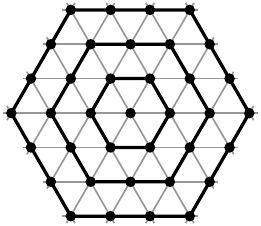}
		\caption{Sets of vertices at the distances one, two and three from the central vertex.}
		\label{fig:NumberVerticesAtDistanceI}
	\end{minipage}
	\qquad
	\begin{minipage}[b]{.58\textwidth}
		\centering
		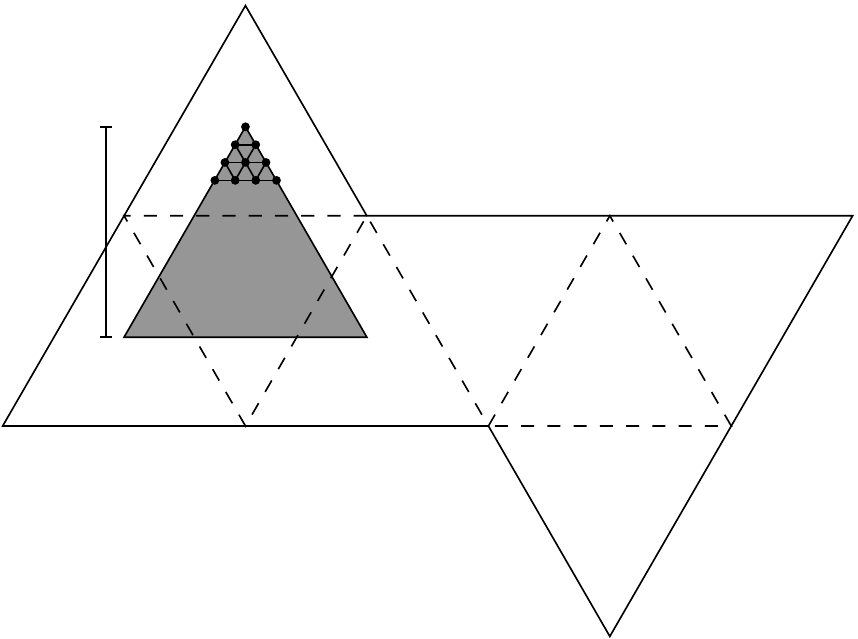
		\caption{Some vertices in $B_{tj}$ with distance of at most $2^j\le n$ from $t\in V$ in an octahedron folding.}
		\label{fig:BoundOfSizeOfABall}
	\end{minipage}
\end{figure}

Theorem~\ref{thm:greedyRoutingAlgorithmFindsSmallPaths} proofs that this routing algorithm finds paths with squared logarithmic length in $n$.
The proof strategy is inspired by the work of Kleinberg~\cite{Kleinberg:2000}.
Lemma~\ref{lem:upperBoundNormalizingFactor} proofs the upper bound of $Z_u$ for all $u\in V$, which is used throughout Section~\ref{sec:c3NotInTheNOctahedralGraph}.
Lemma~\ref{lem:lowerBoundNormalizingFactor} proofs the lower bound of $Z_u$ for all $u$, which is used in the proof of Theorem~\ref{thm:greedyRoutingAlgorithmFindsSmallPaths}.
Let $P_{ui}=\{w\in V|d_{uw}=i\}$ for distances $i\ge1$.
Figure~\ref{fig:NumberVerticesAtDistanceI} shows the vertices in $P_{u1}$, $P_{u2}$ and $P_{u3}$ in the bold cycles, where $u$ is the central vertex.

\begin{lem}
	$Z_u<\ln^{-1}(n+1)$.
	\label{lem:upperBoundNormalizingFactor}
\end{lem}
\begin{pf}
	The longest distance in $G'_n$ is at most $2n$.
	Moreover, $|P_{ui}|\ge i+1>i$ for $i\le n$, as shown in the bold area of the Figure~\ref{fig:BoundOfSizeOfABall}, where $u=t$.
	As $\sum_{i=1}^{n}i^{-1}>\ln(n+1)$, then $$Z_u=\left(\sum_{w\in V\setminus\{u\}}d_{uw}^{-2}\right)^{-1}=\left(\sum_{i=1}^{2n}|P_{ui}|\cdot i^{-2}\right)^{-1}<\left(\sum_{i=1}^n|P_{ui}|\cdot i^{-2}\right)^{-1}<\left(\sum_{i=1}^n i^{-1}\right)^{-1}<\ln^{-1}(n+1).$$\qed
\end{pf}

\begin{lem}
	$Z_u\ge(6\ln(2\textrm{\emph{e}}n))^{-1}$.
	\label{lem:lowerBoundNormalizingFactor}
\end{lem}
\begin{pf}
	As $|P_{ui}|\le6i$ and $\sum_{i=1}^{m}i^{-1}\le\ln(m)+1$, then $$Z_u=\left(\sum_{i=1}^{2n}|P_{ui}|\cdot i^{-2}\right)^{-1}\ge\left(\sum_{i=1}^{2n}6i\cdot i^{-2}\right)^{-1}\ge(6(\ln(2n)+1))^{-1}=(6\ln(2\textrm{e}n))^{-1}.$$\qed
\end{pf}

\begin{thm}
	The greedy routing algorithm performs $\mathcal{O}\left(\log^2n\right)$ expected number of forwards.
	\label{thm:greedyRoutingAlgorithmFindsSmallPaths}
\end{thm}
\begin{pf}
	We partition the path from source $s\in V$ to target $t\in V$ that the greedy routing algorithm finds, such that the algorithm is in phase $j\ge1$ if $2^j<d_{ut}\le 2^{j+1}$, where $u\in V$ is a vertex of the path, and is in phase zero when $d_{ut}\le2$.
	The phase $j$ ends if the current vertex $u$ with the message has a long-range edge to any vertex $v\in B_{tj}$, where $B_{tj}=\left\{w\in V|d_{wt}\le2^j\right\}$.
	By Definition~\ref{dfn:OSW} and Lemma~\ref{lem:lowerBoundNormalizingFactor}, $\Pr(C_{uv})=Z_u\cdot d_{uv}^{-2}\ge\left(6\ln(2\textrm{e}n)d_{uv}^2\right)^{-1}$.
	As there are at least $\sum_{j=1}^{2^j}j$ vertices $w\in V$ such that $d_{wt}\le2^j$ (in the bold area in Figure~\ref{fig:BoundOfSizeOfABall}), then $\left|B_{tj}\right|>2^{2j-1}$.
	Besides, $d_{uv}\le2^{j+1}+2^j<2^{j+2}$ for all $u$ and $v$.
	As all $C_{uv}$ are disjoint, then the probability of the phase $j$ ends in $u$ is $$\Pr\left(\bigcup_{v\in B_{tj}}C_{uv}\right)=\sum_{v\in B_{tj}}\Pr(C_{uv})>\frac{2^{2j-1}}{6\ln(2\textrm{e}n)(2^{j+2})^2}=(192\ln(2\textrm{e}n))^{-1}.$$
	Let $X_j$ be the random variables that count the number of forwards in the phase $j\ge1$.
	As $X_j$ are geometric random variables, then $\E[X_j]<192\ln(2\textrm{e}n)$.
	Let $X$ be the random variable that counts the total number of forwards in the greedy routing.
	As $d_{st}\le2n$, then $X\le2+\sum_{j=1}^{\lceil\log n\rceil+1}X_j$ and $\E[X]<(\lceil\log n\rceil+1)(192\ln(2\textrm{e}n))+2$ by the linearity of expectation.\qed
\end{pf}
\section{C3's not in the n-Octahedral Graph}
\label{sec:c3NotInTheNOctahedralGraph}

The long-range edges generation in \OSW may create new C3's that do not belong to the base $n$-octahedral graph $G'_n$.
The $n$-octahedral graph is a well structured arrangement of C3's and the long-range edges ``hide'' it in $G_n$.
A C3 in $G_n$ and not in $G'_n$ has at least one long-range edge.
Besides, a C3 is a sequence of three edges where a long-range edge may assume any position.
Considering these, the fact that $G_n$ is directed and $u\in V$ is the first vertex of the C3, there are seven distinct compositions of edges that define C3's rooted in $u$, in $G_n$ and not in $G'_n$.
Let $s$ be a directed edge in $G'_n$ and $w$ be a long-range edge generated in \OSW.
These compositions are represented by the events $E_{iu}$.
All events $E_{iu}$ refer to the existence of at least one C3, rooted in $u$, but each specific event having an edge sequence as follows: $E_{1u}$ with edge sequence $(s,s,w)$; $E_{2u}$ with edge sequence $(s,w,s)$; $E_{3u}$ with edge sequence $(s,w,w)$; $E_{4u}$ with edge sequence $(w,s,s)$; $E_{5u}$ with edge sequence $(w,s,w)$; $E_{6u}$ with edge sequence $(w,w,s)$;	$E_{7u}$ with edge sequence $(w,w,w)$.

Then, the event of the existence of at least one C3 in $G_n$, not in $G'_n$ and rooted in $u$ is $E_u=\bigcup_{i=1}^{7}E_{iu}$.
Lemmas from \ref{lem:E1u} to \ref{lem:E7u} bound the probability of each $E_{iu}$ and Theorems \ref{thm:Eu} and \ref{thm:expectedNumberOfC3sNotInTheNOctahedralGraph} proof, respectively, that $\Pr(E_u)$ is $\mathcal{O}\left(\log^{-1}n\right)$ and that the expected number of C3's in $G_n$ and not in $G'_n$ is $\mathcal{O}\left(n^2/\log n\right)$.

\begin{lem}
	$\Pr(E_{1u})<3\ln^{-1}(n+1)$.
	\label{lem:E1u}
\end{lem}
\begin{pf}
	Let $A=\{w\in V|d_{uw}=2\}$.
	Note that $E_{1u}=\bigcup_{a\in A}C_{au}$.
	Using union bound, Definition~\ref{dfn:OSW}, Lemma~\ref{lem:upperBoundNormalizingFactor} and the facts that $d_{au}=2$ for all $a\in A$ and $|A|\le6\cdot2=12$, then $$\Pr(E_{1u})\le\sum_{a\in A}\Pr(C_{au})\le\sum_{a\in A}Z_a\cdot d_{au}^{-2}<12\cdot(\ln(n+1))^{-1}\cdot 2^{-2}=3\ln^{-1}(n+1).$$\qed
\end{pf}

\begin{lem}
	$\Pr(E_{2u})<9/2\ln^{-1}(n+1)$.
	\label{lem:E2u}
\end{lem}
\begin{pf}
	Let $A=\{w\in V|d_{uw}=1\}$ and $B_a=\{w\in V|d_{aw}=1\}$ for all $a\in A$.
	As \OSW does not generate parallel edges when $(u,v)\in E'$ and $C_{uv}$ happens, then $E_{2u}=\bigcup_{a\in A}\bigcup_{b\in A\setminus(\{a\}\cup B_a)}C_{ab}.$
	Using union bound, Definition~\ref{dfn:OSW}, Lemma~\ref{lem:upperBoundNormalizingFactor} and the facts that $d_{ab}=2$, $|A|\le6$ and $|A\setminus(\{a\}\cup B_a)|=|A|-3\le3$ for all $a$ and $b\in A\setminus(\{a\}\cup B_a)$, then $$\Pr(E_{2u})\le\sum_{a\in A}\sum_{b\in A\setminus(\{a\}\cup B_a)}Z_a\cdot d_{ab}^{-2}<6\cdot3\cdot(\ln(n+1))^{-1}\cdot 2^{-2}=9/2\ln^{-1}(n+1).$$\qed
\end{pf}

\begin{lem}
	$\Pr(E_{3u})<36\zeta(3)\ln^{-2}(n+1)$.
	\label{lem:E3u}
\end{lem}
\begin{pf}
	Let $A=\{w\in V|d_{uw}=1\}$ and $B_a=\{w\in V|d_{aw}=1\}$ for all $a\in A$.
	Note that $E_{3u}=\bigcup\limits_{a\in A}\bigcup\limits_{b\in V\setminus(A\cup B_a)}(C_{ab}\cap C_{bu})$.
	Using union bound, $\Pr(E_{3u})\le\sum\limits_{a\in A}\sum\limits_{b\in V\setminus(A\cup B_a)}\Pr(C_{ab}\cap C_{bu})$.
	As $C_{ab}$ and $C_{bu}$ are mutually independent events, so $\Pr(C_{ab}\cap C_{bu})=\Pr(C_{ab})\cdot\Pr(C_{bu})$ for all $a$ and $b\in V\setminus(A\cup B_a)$.
	By this fact, using Definition~\ref{dfn:OSW} and grouping the terms with the same value of $d_{bu}$, $$\Pr(E_{3u})\le\sum_{a\in A}\sum_{b\in V\setminus(A\cup B_a)}Z_a d_{ab}^{-2}\cdot Z_b d_{bu}^{-2}=\sum_{a\in A}\sum_{i=2}^{2n}\sum_{\substack{b\in V\setminus(A\cup B_a)\\d_{bu}=i}}Z_aZ_b \cdot d_{ab}^{-2}d_{bu}^{-2}.$$
	Using Lemma~\ref{lem:upperBoundNormalizingFactor} and the facts that $d_{ab}\ge d_{bu}-1$ and $|V\setminus(A\cup B_a):d_{bu}=i|\le6i$ for all $a$, $b$ and $i$, $$\Pr(E_{3u})<\ln^{-2}(n+1)\sum_{a\in A}\sum_{i=2}^{2n}\sum_{\substack{b\in V\setminus(A\cup B_a)\\d_{bu}=i}}(i-1)^{-2}\cdot i^{-2}\le6\ln^{-2}(n+1)\sum_{a\in A}\sum_{i=2}^{2n}(i-1)^{-2}\cdot i^{-1}.$$
	The proof follows because $i^{-1}<(i-1)^{-1}$ for $i\ge2$, $\sum_{i=2}^{2n}(i-1)^{-3}<\sum_{i=1}^\infty i^{-3}=\zeta(3)$\footnote{\emph{Riemann zeta function} with parameter three is a convergent series such that $\zeta(3)=\sum_{i=1}^\infty i^{-3}<1.20206$.} and $|A|\le6$.\qed
\end{pf}

\begin{lem}
	$\Pr(E_{4u})<3\ln^{-1}(n+1)$.
	\label{lem:E4u}
\end{lem}
\begin{pf}
	Let $A=\{w\in V|d_{uw}=2\}$.
	Note that $E_{4u}=\bigcup_{a\in A}C_{ua}$.
	The proof follows similarly to the Lemma~\ref{lem:E1u} proof, using the fact that $d_{ua}=d_{au}$ for all $a\in A$.\qed
\end{pf}

\begin{lem}
	$\Pr(E_{5u})<36\zeta(3)\ln^{-2}(n+1)$.
	\label{lem:E5u}
\end{lem}
\begin{pf}
	Let $A=\{w\in V|d_{uw}=1\}$ and $B_a=\{w\in V|d_{aw}=1\}$ for all $a\in A$.
	Note that $E_{5u}=\bigcup\limits_{a\in V\setminus(\{u\}\cup A)}\bigcup\limits_{b\in B_a\setminus A}(C_{ua}\cap C_{bu})$.
	In a similar way of the beginning of the Lemma~\ref{lem:E3u} proof and grouping the terms with the same value of $d_{ua}$, $$\Pr(E_{5u})\le\sum_{i=2}^{2n}\sum_{\substack{a\in V\setminus(\{u\}\cup A)\\d_{ua}=i}}\sum_{b\in B_a\setminus A}Z_uZ_b\cdot d_{ua}^{-2}d_{bu}^{-2}.$$
	The proof follows using Lemma~\ref{lem:upperBoundNormalizingFactor} and the facts that $d_{bu}\ge d_{ua}-1$, $|{V\setminus(\{u\}\cup A):d_{ua}=i}|\le6i$ and $|B_a|\le6$ for all $a$, $b\in B_a$ and $2\le i\le2n$.\qed
\end{pf}

\begin{lem}
	$\Pr(E_{6u})<36\zeta(3)\ln^{-2}(n+1)$.
	\label{lem:E6u}
\end{lem}
\begin{pf}
	Let $A=\{w\in V|d_{uw}=1\}$ and $B_a=\{w\in V|d_{aw}=1\}$ for all $a\in A$.
	Note that $E_{6u}=\bigcup\limits_{a\in V\setminus(\{u\}\cup A)}\bigcup\limits_{b\in A\setminus B_a}(C_{ua}\cap C_{ab})$.
	The proof follows similarly to the Lemma~\ref{lem:E5u} proof.\qed
\end{pf}

\begin{lem}
	$\Pr(E_{7u})<36(3\zeta(3)+1/8)\cdot\ln(2n)\cdot\ln^{-3}(n+1)$.
	\label{lem:E7u}
\end{lem}
\begin{pf}
	Let $A=\{w\in V|d_{uw}=1\}$ and $B_a=\{w\in V|d_{aw}=1\}$ for all $a\in A$.
	Note that $E_{7u}=\bigcup\limits_{a\in V\setminus(\{u\}\cup A)}\bigcup\limits_{b\in V\setminus(\{u,a\}\cup A\cup B_a)}(C_{ua}\cap C_{ab}\cap C_{bu})$.
	Using union bound, the fact that the events $C_{ua}$, $C_{ab}$ and $C_{bu}$ are mutually independent, Definition~\ref{dfn:OSW} and Lemma~\ref{lem:upperBoundNormalizingFactor}, $$\Pr(E_{7u})\le\ln^{-3}(n+1)\sum_{a\in V\setminus(\{u\}\cup A)}d_{ua}^{-2}\sum_{b\in V\setminus(\{u,a\}\cup A\cup B_a)}d_{ab}^{-2}d_{bu}^{-2}.$$
	We claim that (i) $\sum\limits_{b\in V\setminus(\{u,a\}\cup A\cup B_a)}d_{ab}^{-2}d_{bu}^{-2}<6(3\zeta(3)+1/8)$ and (ii) $\sum\limits_{a\in V\setminus(\{u\}\cup A)}d_{ua}^{-2}\le6\ln(2n)$.

	For (i), we split the sum in three others: for $d_{ab}<d_{ua}$, $d_{ab}=d_{ua}$ and $d_{ab}>d_{ua}$.
	When $d_{ab}<d_{ua}$, the value of $d_{ua}-d_{ab}$ is positive and the triangle inequality can be used such that $\sum\limits_{b\in V\setminus(\{u,a\}\cup A\cup B_a)}d_{ab}^{-2}d_{bu}^{-2}\le\sum\limits_{b\in V\setminus(\{u,a\}\cup A\cup B_a)}d_{ab}^{-2}(d_{ua}-d_{ab})^{-2}$.
	We group the terms of the sum with the same value of $d_{ab}$ and use the fact that $|V\setminus(\{u,a\}\cup A\cup B_a):d_{ab}=i|\le6i$ such that $$\sum_{b\in V\setminus(\{u,a\}\cup A\cup B_a)}d_{ab}^{-2}d_{bu}^{-2}\le\sum_{i=2}^{d_{ua}-1}\sum_{\substack{b\in V\setminus(\{u,a\}\cup A\cup B_a)\\d_{ab}=i}}i^{-2}(d_{ua}-i)^{-2}\le6\sum_{i=2}^{d_{ua}-1}i^{-1}(d_{ua}-i)^{-2}.$$
	Rearranging the terms, $$\sum_{i=2}^{d_{ua}-1}i^{-1}(d_{ua}-i)^{-2}=\left(\sum_{i=2}^{\left\lceil\frac{d_{ua}-1}{2}\right\rceil}i^{-1}(d_{ua}-i)^{-2}+\sum_{i=1}^{\left\lfloor\frac{d_{ua}-1}{2}\right\rfloor}(d_{ua}-i)^{-1}i^{-2}\right),$$ where the first sum in the parentheses is $0$ for $d_{ua}=3$.
	In both sums, $d_{ua}-i\ge i$ and, when $d_{ab}<d_{ua}$, $$\sum_{b\in V\setminus(\{u,a\}\cup A\cup B_a)}d_{ab}^{-2}d_{bu}^{-2}<6\left(\sum_{i=1}^{\left\lceil\frac{d_{ua}-1}{2}\right\rceil}i^{-3}+\sum_{i=1}^{\left\lfloor\frac{d_{ua}-1}{2}\right\rfloor}i^{-3}\right)<6\left(2\sum_{i=1}^\infty i^{-3}\right)=6\cdot2\zeta(3).$$
	When $d_{ab}=d_{ua}$, as $d_{bu}\ge2$, $|V\setminus(\{u,a\}\cup A\cup B_a):d_{ab}=d_{ua}|\le6d_{ua}$ and $d_{ua}\ge2$, then $$\sum_{b\in V\setminus(\{u,a\}\cup A\cup B_a)}d_{ab}^{-2}d_{bu}^{-2}\le(1/4)\sum_{b\in V\setminus(\{u,a\}\cup A\cup B_a)}d_{ab}^{-2}\le(1/4)\cdot6d_{ua}\cdot d_{ua}^{-2}\le6/8.$$
	When $d_{ab}>d_{ua}$, we use the triangle inequality, group the terms of the sum with the same value of $d_{ab}$ and use the fact that $|V\setminus(\{u,a\}\cup A\cup B_a):d_{ab}=i|\le6i$ such that $$\sum_{b\in V\setminus(\{u,a\}\cup A\cup B_a)}d_{ab}^{-2}d_{bu}^{-2}\le6\sum_{i=d_{ua}+1}^{2n}i^{-1}(i-d_{ua})^{-2}=6\sum_{i=1}^{2n-d_{ua}}(d_{ua}+i)^{-1}i^{-2}<6\sum_{i=1}^\infty i^{-3}=6\zeta(3).$$
	Therefore, $\sum\limits_{b\in V\setminus(\{u,a\}\cup A\cup B_a)}d_{ab}^{-2}d_{bu}^{-2}<6(3\zeta(3)+1/8)$.

	For (ii), we group the sum terms with the same value of $d_{ua}$ and use the facts that $|V\setminus(\{u\}\cup A):d_{ua}=i|\le6i$ and $\sum_{i=1}^{m}i^{-1}\le\ln(m)+1$ such that $$\sum_{a\in V\setminus(\{u\}\cup A)}d_{ua}^{-2}=\sum_{i=2}^{2n}\sum_{\substack{a\in V\setminus(\{u\}\cup A)\\d_{ua}=i}}d_{ua}^{-2}\le6\sum_{i=2}^{2n}i^{-1}\le6(\ln(2n)+1-1)=6\ln(2n).$$\qed
\end{pf}


\begin{thm}
	$\Pr(E_u)$ is $\mathcal{O}\left(\log^{-1}n\right)$.
	\label{thm:Eu}
\end{thm}
\begin{pf}
	Note that $E_u=\bigcup_{i=1}^{7}E_{iu}$ and, using union bound and Lemmas from \ref{lem:E1u} to \ref{lem:E7u}, $$\Pr(E_u)<21/2\ln^{-1}(n+1)+108\zeta(3)\ln^{-2}(n+1)+36(3\zeta(3)+1/8)\cdot\ln(2n)\cdot\ln^{-3}(n+1).$$
	As $\ln(2n)/\ln(n+1)$ is $\mathcal{O}(1)$, then $\Pr(E_u)$ is $\mathcal{O}\left(\log^{-1}n\right)$.\qed
\end{pf}

\begin{thm}
	The expected number of C3's in $G_n$ and not in $G'_n$ is $\mathcal{O}\left(n^2/\log n\right)$.
	\label{thm:expectedNumberOfC3sNotInTheNOctahedralGraph}
\end{thm}
\begin{pf}
	Let $\mathcal{C}_u$ be the sets of all C3's that contain $u$ and have at least one long-range edge, for all $u\in V$.
	Let $\mathcal{C}_{iu}$ be the partitions of $\mathcal{C}_u$ with all C3's that have the same edge sequence defined in the event $E_{iu}$, for all $1\le i\le7$ and $u$.
	Note that the cycles in $\mathcal{C}_{iu}$ are sequences of three vertices such that $u$ is the first vertex in all.
	Let $X_u$ be the random variables that count the number of C3's that contain $u$ and have at least one long-range edge, for all $u$.
	Let $X_c$ be the Bernoulli random variables that are one if the C3 $c$ is in $G_n$ or $0$ otherwise, for all $c\in\mathcal{C}_u$.
	Note that $X_u$ and $X_c$ are distinct random variables.
	Given the definitions, $X_u=\sum_{c\in\mathcal{C}_u}X_c=\sum_{i=1}^7\sum_{c\in\mathcal{C}_{iu}}X_c$.

	We compute $\E\left[\sum_{c\in\mathcal{C}_{iu}}X_c\right]$ for all $1\le i\le7$ in a similar way of the sequence of Lemmas from \ref{lem:E1u} to \ref{lem:E7u}.
	For $i=1$, the value of $\E\left[\sum_{c\in\mathcal{C}_{1u}}X_c\right]=\sum_{c\in\mathcal{C}_{1u}}\Pr\left(X_c=1\right)$, by linearity of expectation and the fact that $X_c$ are Bernoulli random variables.
	We enumerate the cycles in $\mathcal{C}_{1u}$ in a similar way of the Lemma~\ref{lem:E1u} proof.
	Let $A=\{v\in V|d_{uv}=2\}$ and $B_w=\{v\in V|d_{wv}=1\}$ for $w\in V$, then $$\E\left[\sum_{c\in\mathcal{C}_{1u}}X_c\right]=\sum_{a\in A}\sum_{b\in(B_u\cap B_a)}\Pr\left(C_{au}\right)\le2\sum_{a\in A}\Pr\left(C_{au}\right)<6\ln^{-1}(n+1),$$ because $|B_u\cap B_a|\le2$ and $\sum_{a\in A}\Pr\left(C_{au}\right)<3\ln^{-1}(n+1)$, as Lemma~\ref{lem:E1u} shows.
	In fact, obtaining the upper bounds of the next $\E\left[\sum_{c\in\mathcal{C}_{iu}}X_c\right]$, for $2\le i\le7$, follows similarly to Lemmas from \ref{lem:E2u} to \ref{lem:E7u}, except for $i=4$, which follows similarly to $i=1$.
	Let $X$ be the random variable that counts the number of C3's with at least one long-range edge.
	Therefore, $X=1/3\sum_{u\in V}X_u=1/3\sum_{u\in V}\sum_{i=1}^7\sum_{c\in\mathcal{C}_{iu}}X_c$, because each $c$ is considered three times in distinct $u$'s.
	By linearity of expectation, $\E[X]=1/3\sum_{u\in V}\sum_{i=1}^7\E\left[\sum_{c\in\mathcal{C}_{iu}}X_c\right]$ and, then $\E[X]<\lambda\left(n^2/\log n\right)$ by Lemma~\ref{lem:numberOfVertices}.\qed
\end{pf}

As the size of $G'_n$ is $\Theta\left(n^2\right)$, by Lemmas \ref{lem:numberOfVertices} and \ref{lem:numberOfEdges}, and the number of long-range edges in $G_n$ are at most $|V|$, by Definition~\ref{dfn:OSW}, then the size of $G_n$ is $\Theta\left(n^2\right)$.
So, Theorem~\ref{thm:expectedNumberOfC3sNotInTheNOctahedralGraph} also shows that the expected number of C3's in $G_n$ and not in $G'_n$ is sublinear in the size of $G_n$.
Using Euler characteristic
and Lemmas \ref{lem:numberOfVertices} and \ref{lem:numberOfEdges}, the number of C3's in $G'_n$ is $8n^2$.
This implies in the Corollary~\ref{cor:numberOfC3sInAnOSWGraph} statement.

\begin{cor}
	The expected number of C3's in an \OSW graph $G_n$ is $\Theta\left(n^2\right)$.
	\label{cor:numberOfC3sInAnOSWGraph}
\end{cor}
\section{Conclusion}
\label{sec:conclusion}

\OSW model generates small world graphs with size $\Theta\left(n^2\right)$ and has a greedy routing algorithm that performs $\mathcal{O}\left(\log^2n\right)$ expected forwards. The base graph can be embedded in spheres preserving constant bounded distances.
It generates long-range edges with the inverse $2^\textrm{th}$-power distribution, this implies in the routing algorithm finding small paths.
\OSW graphs have expected linear number of C3's, allowing the exhaustive searching of C3's for any desired purpose.
We suggest for future works the design of an algorithm that finds the $n$-octahedral graph running searches of C3's in each vertex.
\section*{Acknowledgments}

We thank the Coordination for the Improvement of Higher Education Personnel (CAPES).

\section*{References}

\bibliography{bibliography}

\end{document}